\documentclass[10pt,a4paper]{iopart}

\usepackage{amssymb}
\usepackage{wrapfig}
\usepackage{graphicx}
\usepackage[usenames]{color}
\usepackage[citestyle=numeric]{}

\begin{document}

\title[Buffer-gas cooling of molecules in the low-density regime]{Buffer-gas cooling of molecules in the low-density regime: Comparison between simulation and experiment}

\author{Thomas Gantner}
\author{Manuel Koller}%
\author{Xing Wu*}%
\author{Gerhard Rempe}%
\author{Martin Zeppenfeld}%
\ead{martin.zeppenfeld@mpq.mpg.de}
{Max Planck Institute of Quantum Optics, Hans-Kopfermann-Str. 1, 85748 Garching, Germany}\\
\**Current address: {Department of Physics, Yale University, New Haven, Connecticut 06511, USA; Department of Physics, Harvard University, Cambridge, Massachusetts 02138, USA}

\date{\today}

\begin{abstract}
Cryogenic buffer gas cells have been a workhorse for the  cooling of molecules in the last decades. The straightforward sympathetic cooling principle makes them applicable to a huge variety of different species. Notwithstanding this success, detailed simulations of buffer gas cells are rare, and have never been compared to experimental data in the regime of low to intermediate buffer gas densities. Here, we present a numerical approach based on a trajectory analysis, with molecules performing a random walk in the cell due to collisions with a homogeneous buffer gas. This method can reproduce experimental flux and velocity distributions of molecules emerging from the buffer gas cell for varying buffer gas densities. This includes the strong decrease in molecule output from the cell for increasing buffer gas density and the so-called boosting effect, when molecules are accelerated by buffer-gas atoms after leaving the cell. The simulations provide various insights which could substantially improve buffer-gas cell design.
\end{abstract}

\maketitle



\section{Introduction}
Cooling molecules in a cryogenic buffer gas cell is a universal technique to produce cold and relatively slow beams  of molecules \cite{Hutzler2011}. As it does not depend on molecular properties like internal energy structure or electric dipole moment, it has been applied to a huge variety of species within the last two decades. This ranges from small, light and chemically stable polyatomic molecules like $ND_3$ \cite{vanBuuren2009,Patterson2009} or heavy species like $ThO$ \cite{Vutha2010}, to radicals such as $NH$ \cite{Egorov2004}, $CaF$ \cite{Truppe2018}, or $SrF$ \cite{Barry2011}, to big biomolecules like $trans-cinnamaldehyde$ \cite{Zinn2015}, just to name a few. This method brought big advancements for various fields of physics, such as astrophysics \cite{Mengel2000}, biophysics \cite{Patterson2013} or the quest for more and more precise determination of fundamental physical constants \cite{Tarbutt2013, ACMECollaboration2018, Hudson2006b}. 

Notwithstanding this success, detailed simulations of buffer gas cells are rare. Simulations of the high buffer gas density regime, when the mean free path of a molecule is short compared to all typical dimensions of the buffer gas cell were performed \cite{Bulleid2013,Singh2018}. Simulations of the low and intermediate density regime were performed by Doppelbauer et al. \cite{Doppelbauer2017}. However, due to a high computational cost this was limited to a specific buffer gas density and a cell length of $3\,$mm. Therefore, the results of the simulation could not be compared to experimental data. This density regime is particularly important since it is necessary to produce a molecule beam which is not only internally cold, but also slow in the lab frame.
 
Here, we present a simplified approach to modeling the low-density buffer-gas cell environment, with Knudsen numbers $K_n$ ranging from above 10 to about 0.03 in the cell. $K_n$ is defined as the ratio of the mean free path $l$ of a molecule to the characteristic length of a system $l_{char}$, $K_n=l/l_{char}$ \cite{Demtroder2002}. Molecules perform a random walk inside the cell due to repeated collisions with buffer-gas atoms, captured by a simple toy-model for collision kinetics. The results of the numerical simulation exceed expectations, and provides excellent agreement between numerical and empirical data with only one free fit parameter. In particular, it reproduces quantitatively the strong decrease in molecule signal from the buffer-gas cell for increasing buffer-gas density. An intuitive understanding of this phenomenon is provided by an analogy to the Gambler's ruin problem \cite{Coolidge1909}. With a few additional refinements the simulation can even be brought to fit empirically measured velocity distributions of the molecules. This includes the boosting effect \cite{Motsch2009} that leads to the extinction of the slowest molecules, as a result of collisions with faster buffer gas atoms at the vicinity of the cell exit nozzle. This effect appears even at moderate buffer gas densities with a corresponding Knudsen number down to about $0.1$, where the characteristic size $l_{char}$ equals the nozzle diameter. In the end, we show that the simulations also reveal simple scaling laws for the buffer gas cell performance. These can be used to develop a scheme to identify important parameters for cell optimization, and suggest an improvement to our current setup, which might substantially increase the flux of slow molecules.  Additionally, this method can significantly improve the currently used empirical trial-and-error approach, shortening the development time on cryogenic beam sources of future experiments. 

\begin{figure}
\centering
  \includegraphics[width=1\textwidth]{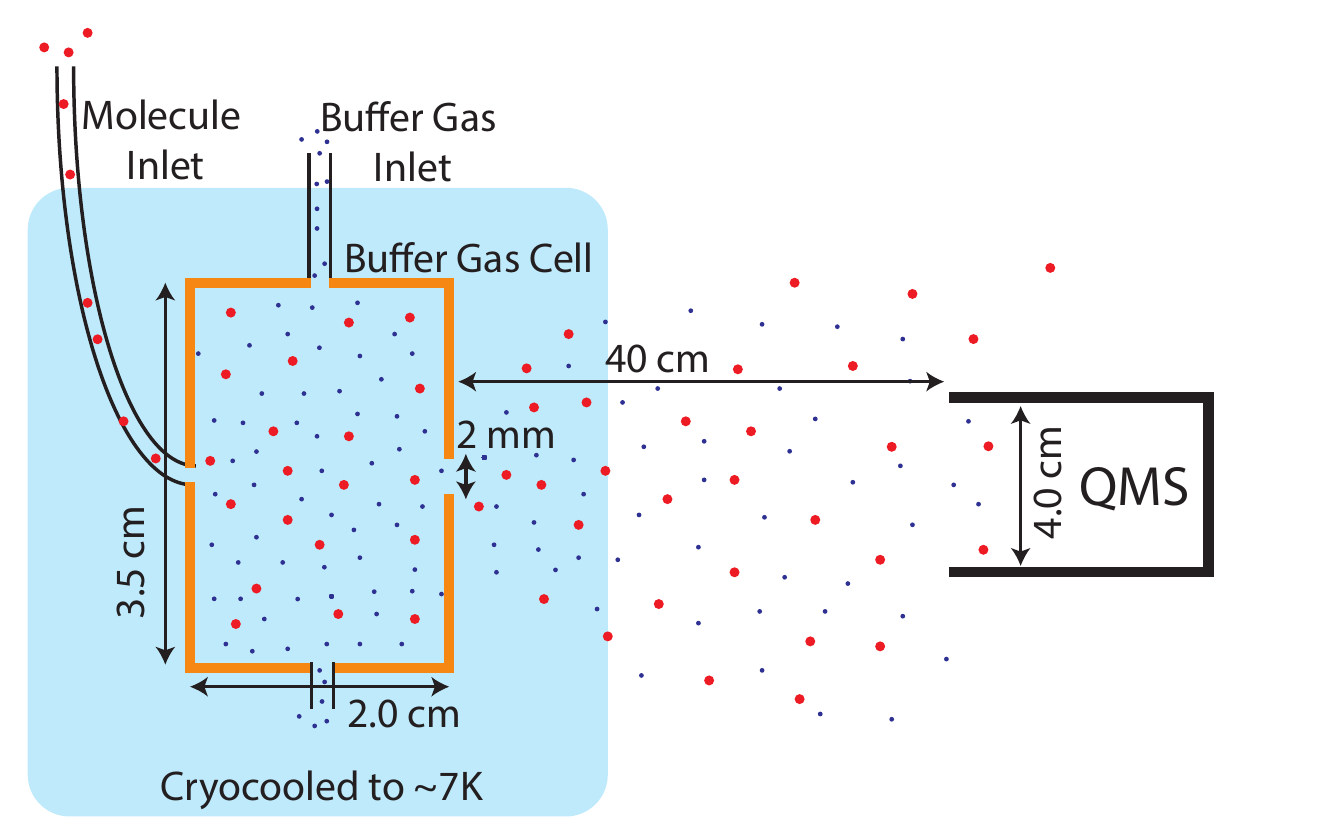}
  \caption{Sketch of the experimental setup. Molecules (depicted as red dots) are brought into the buffer gas cell by a heated tube. Precooled buffer gas atoms (depicted as blue dots) are flown into the cell through the cylindrical wall. Molecules and buffer gas emerge through the nozzle opposite to the molecule inlet hole, and are detected after traveling 40cm through the vacuum chamber by a quadrupole mass spectrometer. The cryocooled environment of the buffer gas cell acts as a vacuum pump.}
  \label{fig:Setup}
\end{figure}

\section{Simulation method}
\subsection{Setup geometry}
The simulations are based on the experimental setup which is shown in figure \ref{fig:Setup}. This setup has been used with slight variations for previous experiments using buffer gas cooled molecules \cite{Wu2016, Wu2017}. The molecules are transported through the back side of the cell via a Teflon tube of $1\,$mm inner diameter. This molecule inlet is thermally isolated from the cell, and stabilized to a temperature to prevent freezing of molecules inside. This temperature will be the starting temperature for molecules in the simulation. The buffer gas feed line in contrast is precooled to the cell temperature, to be able to maintain a low cell temperature of down to about $7\,$K. The buffer gas is delivered by eight symmetrically arranged holes perpendicular to the molecule input, to obtain a homogenous buffer gas distribution inside the cell.  The cell itself consists of a copper cylinder with a length of $2\,$cm and a diameter of $3.5\,$cm. The output nozzle is placed in the center of the plane opposite to the molecule input hole, with a diameter $D$ of $2\,$mm. Outside the exit nozzle, in a straight line with the molecule input hole and the exit nozzle, the molecules are detected by a quadrupole mass spectrometer (QMS, Pfeiffer Prisma QMG 220) at a distance of $40\,$cm. The effective detection area has a radius of about $2\,$cm. For the velocity distribution measurement, a bent electrostatic quadrupole guide is placed after the cell followed by a straight guide as a time-of-flight stretch, as described previously \cite{vanBuuren2009}. For the velocity distribution measurements, a Pfeiffer QMG 700 is used as molecule detector.

\subsection{Details of the numerical model}
The simulation captures the low-density dynamics by tracing single molecules on their random walk through this buffer gas cell. The molecules interact with the buffer gas atoms via elastic collisions, and propagate ballistically between two collision events. As a necessary simplification, all molecule-molecule collision effects are ignored. The trajectories of the molecules start in the center of the molecule inlet, randomly picking a starting velocity from a Maxwell-Boltzmann distribution according to the temperature of the molecule input tube. From the inlet, taken as a point-like source, the molecules propagate in a straight line, with the propagation direction randomly chosen from a solid angle of $2\pi$. The distance of free propagation $d$ is randomly selected according to a probability following a Beer-Lambert law, $p(d)\propto e^{-d/l}$, where $l$ is the mean free path. After the free propagation, a molecule-buffer-gas-atom collision is simulated. For this collision, a buffer gas atom velocity vector is selected from a Maxwell-Boltzmann distribution according to the cell temperature. The molecule's velocity and direction after each collision are computed from momentum and energy conservation laws. One important simplification is that the collision angle in the center of mass frame is randomly assigned, homogeneous in $4\pi$. As many collisions of the molecule with the atoms are necessary for thermalization, a possible error due to the simplified angle distribution averages out. The ballistic flight and elastic collision steps are then repeated until the molecule either reaches the exit nozzle and is transmitted out of the cell, or hits the boundary of the cell. As we assume a $100\%$ sticking probability of the molecules on the cold cell walls, justified by the typically very low vapor pressure of molecules at the temperatures considered, it is then lost from the simulation.

\begin{figure}
\centering
  \includegraphics[trim={0cm 0cm 0cm 0cm},clip,width=1\textwidth]{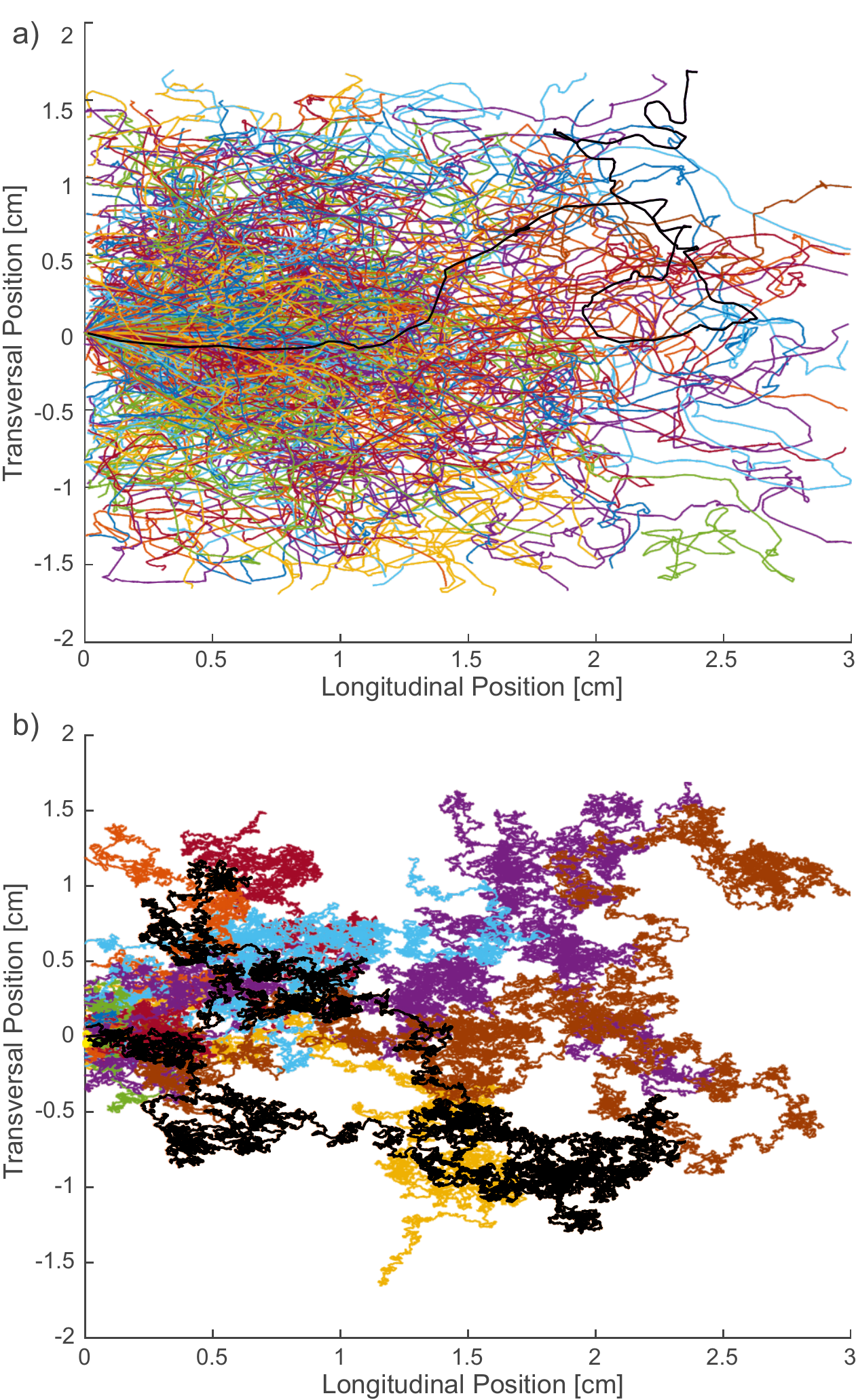}
  \caption{Simulated molecule trajectories inside the buffer gas cell for moderate (a) and high (b) buffer gas densities. One trajectory in each subfigure is shown in black for clarity. In the high density case, a more typical trajectory marked in bright yellow diffuses hardly more than $1\,$mm into the cell and gets lost directly next to the input nozzle. (a) and (b) both show a total of 200 randomly chosen trajectories.
}
  \label{fig:Trajectories}
\end{figure}

Examples of the trajectories resulting from this simulation are shown for two different buffer gas densities in figure \ref{fig:Trajectories}. The molecules start at the origin of the coordinate system, (0,0), and the exit nozzle is at (3,0), in the center of the opposite wall of the cell. For the moderate buffer gas density shown in figure \ref{fig:Trajectories}a), the mean free path of the molecules is still similar compared to the output nozzle, but  much smaller than the cell length. Therefore they travel about $0.5\,$cm into the cell before they are fully thermalized. Afterwards, they diffuse through the cell, until their trajectories terminate at a cell wall. For the trajectories shown in figure \ref{fig:Trajectories}a), one of the molecules could exit the cell, and would be counted as successfully cooled. At higher densities, as shown in figure \ref{fig:Trajectories}b), the mean free path of the molecules is very short, and therefore, they thermalize very close to the entry nozzle. Only few trajectories diffuse further than $0.5\,$cm into the cell. 

For the trajectories shown here, spatial variations of density or temperature of the buffer gas are ignored. This is a key simplification important to maintain short computing times. It is justified by maintaining a low buffer gas density in the cell in the experiment. The gas should therefore be well thermalized, and effects of hydrodynamic flow or pressure gradients should be small enough to be neglected. In the vicinity of the exit nozzle, however, this simplification becomes unrealistic as we maintain vacuum conditions outside the buffer gas cell, and the collision frequency of the molecules is altered. Buffer gas atoms coming from the direction of the output nozzle are excluded in the simulation whenever a molecule-atom collision occurs closer to the nozzle than one mean free path of a helium atom. 

In contrast, just outside the cell the buffer gas atoms can only come from the direction of the nozzle, and all other directions are excluded. This necessarily reduces the collision probability and ultimately leads to a free propagation of the molecules away from the cell. Additionally it results in a net acceleration of the molecules, as buffer gas atoms only come from one general direction. This creates the boosting effect on the molecule's velocity distribution, without needing to take into account the formation of a hydrodynamic buffer gas beam. The simulation ends when free propagation is achieved.

An additional complication of the simulation necessary to reproduce a boosting effect is the variation of the mean free path with the velocity of the molecules. The mean free path of a molecule inside the buffer gas environment changes according to its velocity. For example, even a molecule at standstill would collide with a buffer gas atom eventually due to the motion of the atoms, therefore, its corresponding mean free path would have been zero. Therefore, the mean free path of a molecule is multiplied by a correction factor equal to the ratio of its velocity $v_{\mathrm{mol}}$ to the average relative velocity $\overline{v_{\mathrm{rel}}}(v_{\mathrm{mol}})$ between the molecule and the buffer gas atoms. This yields the refined version of the mean free path, 
\begin{equation}
l^*=\frac{1}{n\sigma}\frac{v_{\rm mol}}{\overline{v_{\rm rel}}}=l \frac{v_{\rm mol}}{\overline{v_{\rm rel}}},
\label{eq:refinedlambda}
\end{equation}
where $l=1/(n\sigma)$ is the simple form of mean free path for buffer-gas density $n$ and molecule-atom collision cross section $\sigma$.This correction was a key step to achieve a good fit between the simulation and the experimentally obtained velocity distributions.

\begin{figure}
\centering
  \includegraphics[trim={0cm 0cm 0cm 0cm},clip,width=1\textwidth]{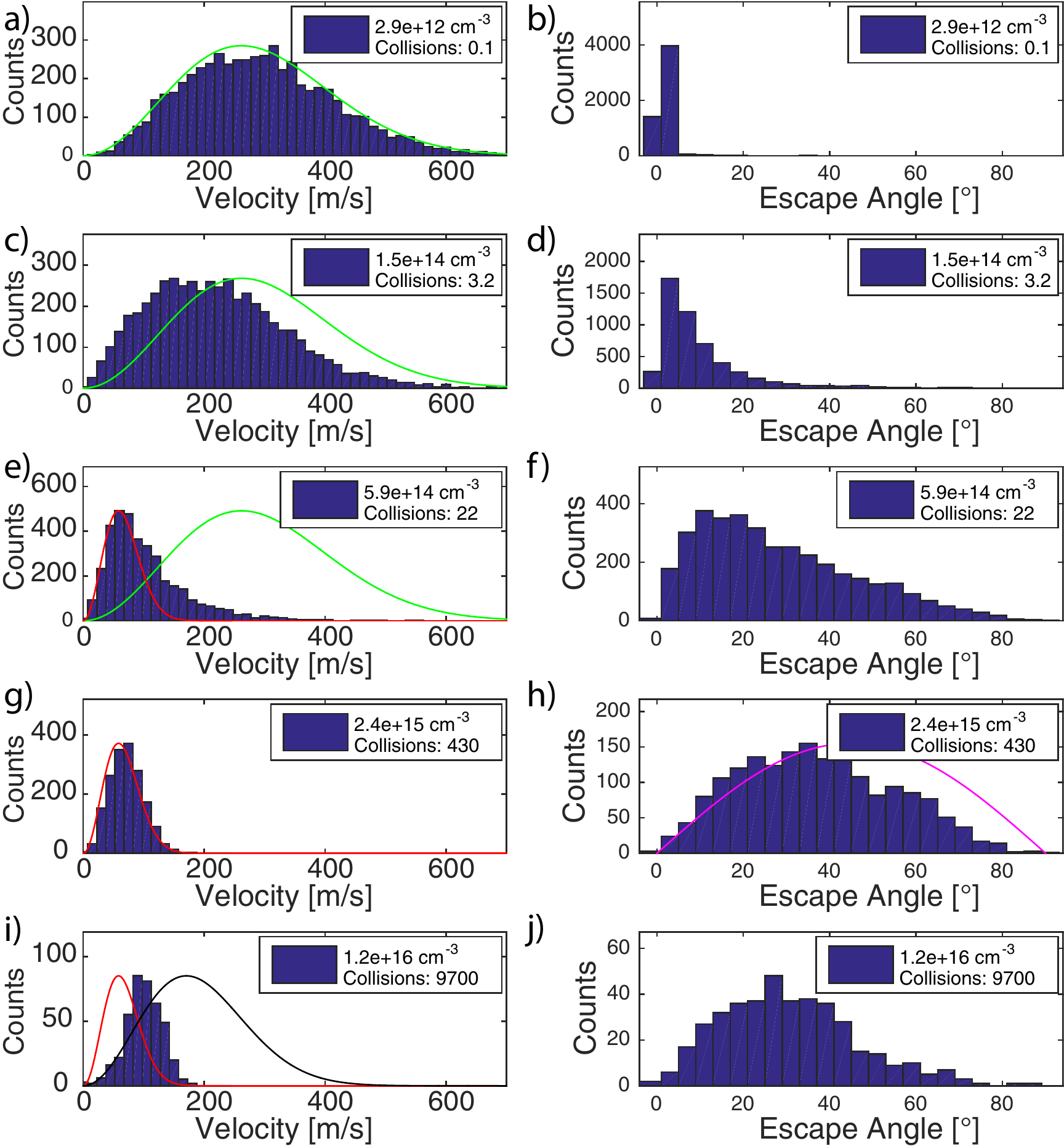}
  \caption{Simulated velocity distributions (a, c, e, g, i) and angle distributions (b, d, f, h, j) of molecules coming out of the buffer gas cell at increasing buffer gas densities and number of collisions per molecule from top to bottom. The green lines in a), c) and e) show the Maxwell-Boltzmann distribution of the molecules at starting temperature. The red lines in e), g) and i) show the Maxwell-Boltzmann distribution of the molecules at cell temperature. The black line in i) shows the Maxwell-Boltzmann distribution of the buffer gas atoms at cell temperature. The distribution $cos(\alpha) \cdot sin(\alpha)$ of the escape angle $\alpha$ which is expected for a purely effusive source is shown as a magenta line in h. The average number of collisions per molecule is in the figure legend. The density in the legend is based on the comparison to experimental data, see figure \ref{fig:FluxComparison}. Further details are given in the text.}
  \label{fig:SimulationOutput}
\end{figure}

\section{Comparing simulations with measurements}
The simulation provides the final velocity vector for each simulated molecule. Figure \ref{fig:SimulationOutput} shows the results for different buffer gas densities and consequently different number of collisions per molecule. \ref{fig:SimulationOutput}a), c), e), g), and i) depict the obtained velocity distributions, and b), d), f), h), and j) the corresponding angular distributions, to confirm the plausibility and qualitative agreement with experiments. To clearly show the cooling effect, the velocity distributions are compared to Maxwell-Boltzmann distributions. These simulations shown here are performed for $CH_3F$ cooled by Helium, with a cell temperature of $7\,$K. In a) and b), the applied buffer gas density is small, and few collisions occur in the cell. Therefore, most molecules that escape the cell are not cooled. As expected, the output velocity distribution closely resembles the input velocity distribution according to a Maxwell-Boltzmann distribution of the uncooled molecules (green curve). The angular distribution shown in b) results from molecules propagating in a straight line from the molecule input hole, spatially filtered at the exit nozzle. Subfigures c) to h) show the results for an increasing buffer gas density, until the molecules are efficiently cooled to the temperature of the buffer gas atoms in g) and h). The velocity distribution narrows, and finally closely matches the Maxwell-Boltzmann distribution of molecules at cell temperature (red curve). The angular distribution widens considerably, and matches closely the expected curve for an effusive source as shown in magenta. Subfigures i) and j) display the onset of boosting for high densities. In i), the red curve shows again a Maxwell-Boltzmann distribution of molecules at cell temperature, whereas the black curve shows the Maxwell-Boltzmann distribution of the buffer gas atoms. While the velocity distribution is still narrower than for uncooled molecules, it is clearly visible that molecules from the thermal distribution are shifted towards the faster velocity of the atoms. This is due to the fact that molecules are hit predominantly by buffer gas atoms with forward velocity in the vicinity of the nozzle. As the buffer gas atoms are much lighter and thus faster, this leads to a net acceleration. This has been observed in different setups and was dubbed boosting \cite{Motsch2009}, and prevented the production of cold and slow molecular beams directly from a buffer gas cell. Larger buffer gas densities have not been studied in these simulations, as flow dynamics would have to be considered, and the assumption of a homogeneous buffer gas density is expected to break down. This would make different simulation methods necessary. These kind of simulations have been performed previously \cite{Bulleid2013,Singh2018,Doppelbauer2017}, and are beyond the scope of this paper.

\subsection{Benchmark cell throughput efficiency}
The main objective for the simulations is to provide understanding of the change of molecule throughput efficiency of the cell depending on the buffer gas density. In the simulations the density is varied by more than 4 orders of magnitude, and thereby the probability of a molecule to reach the output changes by more than 2 orders of magnitude. The simulations are performed for different molecules, $CH_3F$ and $ND_3$ to test different molecule masses and input temperatures. Moreover, the cell temperature is varied and the buffer gas species is alternated between Helium and Neon, thereby changing the collision kinetics. The number of molecules leaving the cryogenic cell through the exit nozzle with a solid angle corresponding to the QMS detection area is recorded. All measurements are performed for low molecule flow rates into the cell ($0.02\,$sccm to $0.05\,$sccm), justifying the assumption of no molecule-molecule collisions in the simulation.

\begin{figure}
\centering
  \includegraphics[width=1\textwidth]{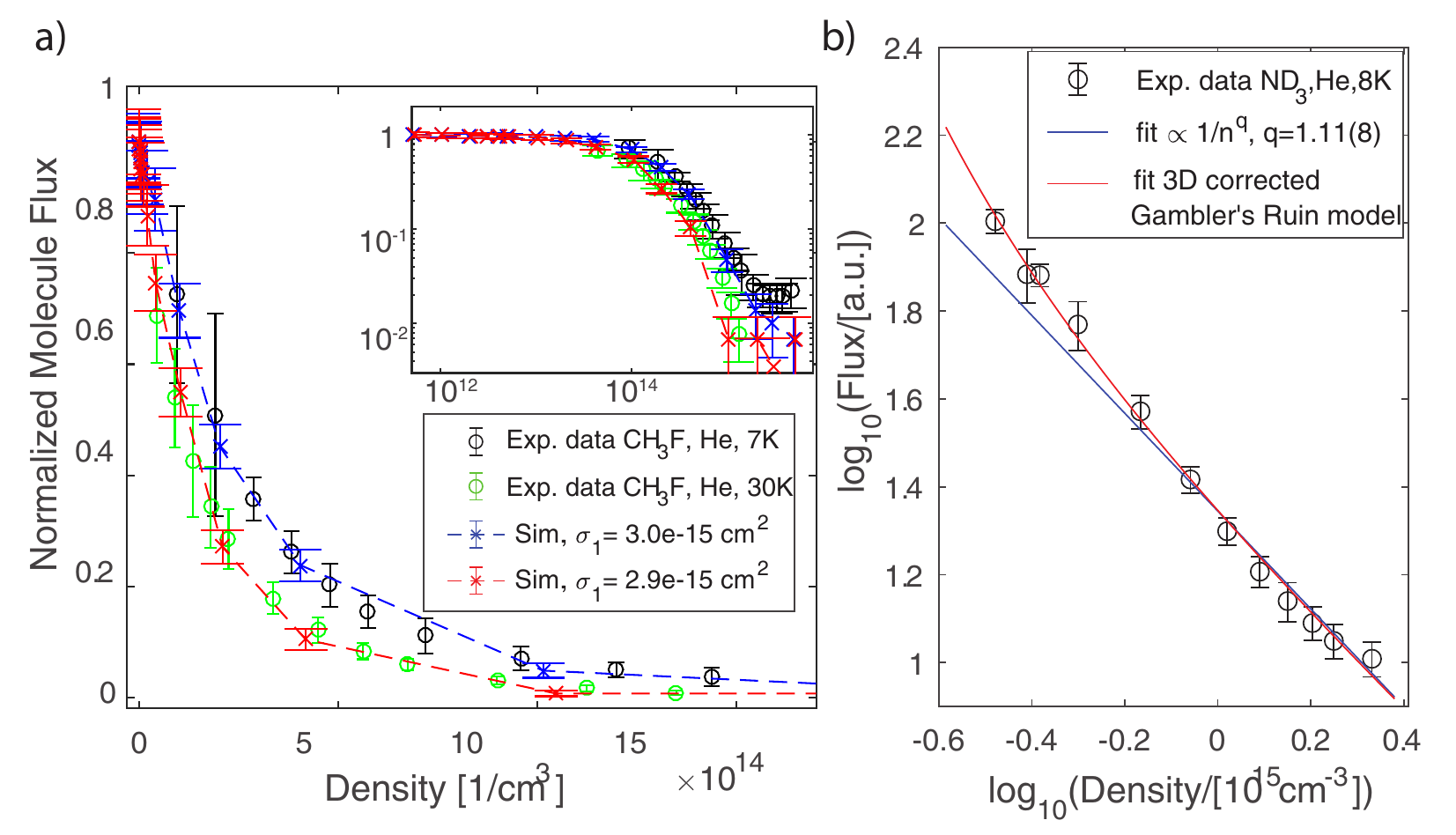}
  \caption{Comparison between simulated (crosses) and experimental (circles) data, for different cell temperatures, $7\,$K (black and blue) and $30\,$K (green and red). Data in a) is plotted in linear scale, while the inset is in double-logarithmic scale. b) Fit the throughput vs. density dependence with a basic power law dependence and with an analytical model based on the Gambler's Ruin problem (including 3D correction). The power-law fit excludes the first 4 data points which deviate from the straight line. The values in the parentheses are the $1\sigma$ statistical error from the fit. The buffer gas flow rate in the experiment ranged from $0.02$ to $1.8\,$sccm}
  \label{fig:FluxComparison}
\end{figure}

We compare the simulation results to the molecule flux measured at the detector. In the experiments, the buffer gas density in the cell is varied by adjusting the flux into the cell. Figure \ref{fig:FluxComparison}a) shows the comparison of the experimental data in green and black and the simulation results in blue and red for $CH_3F$ and $He$, with cell temperatures of $7\,$K and $30\,$K, respectively. The results are scaled vertically to overlap the simulated and measured values at zero buffer gas density, where the cell exit nozzle acts as a pure aperture for the molecules. The throughput efficiency at this point is given by the solid angle of the nozzle. For our cell geometry, this results in an efficiency of about $0.1\%$. To fit in horizontal direction, the effective collision cross section $\sigma$ is varied, adjusting the mean free path of the molecules in the buffer gas environment. In Figure \ref{fig:FluxComparison}a), the same data is plotted in both linear and double-logarithmic (in figure inset) scale.

For small buffer gas densities, the throughput of molecules through the cell towards the QMS is constant. In this regime, the mean free path is longer than the cell length, and molecules are not scattered out from their direct path to the exit nozzle. Small changes of the density  have no influence on the molecules. This corresponds to the no-cooling regime in \ref{fig:SimulationOutput}a) and \ref{fig:SimulationOutput}b). For increasing densities, the molecules are more and more likely to be scattered, and consequently lost at the cell wall, and the molecule throughput decreases. The numerical data reproduces the empirical data in this regime. 

At the moderate buffer gas density regime where molecules undergo substantial thermalization in the cell, the decrease of molecule signal follows roughly a $\propto 1/n$ dependence. In Figure \ref{fig:FluxComparison}b), the experimental data for the cell throughput is fit with both a $1/n^q$ dependence where the extracted power $q$ is approximately $1$, and an analytic form based on the Gambler’s Ruin model \cite{Coolidge1909}. At low buffer gas densities, the molecule would travel a long distance towards the exit before being stopped by collisions, and therefore the diffusion process would start closer to the exit nozzle. This gives the molecule a higher chance to leave the cell. At high densities, in contrast, the molecules thermalize close to the input nozzle, giving them a high chance to diffuse back to the cell wall and freeze there. We formulate this in one dimension as the statistical Gambler’s Ruin problem as follows. We consider a molecule initially at a distance $x=m \cdot l$ from the cell input, where $l=1/n\cdot\sigma$ is the mean free path for buffer-gas density $n$ and molecule-atom collision cross section $\sigma$, and $m$ is the number of collisions necessary to initially thermalize the molecule. From the position $x$, a molecule has an even chance to move a distance $l$ forward or backward, and this process repeats. If the position of the molecule decreases below zero, the molecule is lost, whereas if it increases above the cell length $L$, it has a fixed probability to exit the cell successfully. This is equivalent to a gambler placing a series of even bets, where it is well known that the probability to reach a value above $L$ from a starting value $x$ without previously reaching values below $0$ is $x/L$ \cite{Coolidge1909}. We thus obtain a probability for the molecule to reach the cell exit 
\begin{equation}
P_{\rm 1D}\propto \frac{x}{L}=\frac{m}{n\sigma L}
\label{eq:P1D}
\end{equation}

Additionally, we can include a 3D correction factor. In particular, from distance $x$ onwards, molecules undergo diffusion, so the probability for reaching the cell exit is proportional to the corresponding solid angle,
$\Delta \Omega=\frac{A\cos\theta}{4\pi(L-x)^2/\cos^2\theta}=\frac{D^2\cos^3\theta}{16L^2(1-x/L)^2}$, where $A=\pi(D/2)^2$ is the area of the cell exit hole with diameter $D$, and $\theta$ is the polar angle of a molecule with respect to the exit of the cylindrically symmetric cell. Integrating over the angular dependence results in a 3D corrected probability for a molecule to reach the cell exit,
\begin{equation}
P_{\rm 3D}(n) \propto \int d\phi \int \sin\theta d\theta P_{\rm 1D} \Delta\Omega=\frac{a}{n}\frac{b}{(1-a/n)^2}
\label{eq:P3D}
\end{equation}
with the constants $a=m/(\sigma L)$ and $b=\pi D^2/(8L^2)$. This model holds when there is sufficient thermalization of molecules inside the buffer-gas cell. With increasing density, Equation~\ref{eq:P3D} reduces to approximately the $\propto 1/n$ dependence expected from the 1D Gambler's Ruin model (Equation~\ref{eq:P1D}), as shown in Figure \ref{fig:FluxComparison}b). Hence, the analytical approximation based on  a simple Gambler's Ruin model provides a qualitative understanding of the escape probability of molecules out of a buffer gas cell. 

The numerical simulation, however, gives a better quantitative fit over a larger density range. We want to emphasize that the only free fit parameter in the simulation is the mean free path of the molecules in the buffer gas environment. This shows that in the low to medium density regime, a straightforward cell model, resulting in a random walk of molecules through a homogeneous buffer gas background, works well to predict the propagation of molecules through the cell, and ultimately provides a prediction of the flux of molecules leaving the buffer gas cell.

At high densities, however, both the analytical approximation and the numerical simulation break down. At the highest buffer gas densities considered in figure \ref{fig:FluxComparison}, a deviation between simulation and experiment is visible, most pronounced in the measurement taken at 7K cell temperature. As the buffer gas density is increased more and more, hydrodynamic effects start to play a role. This leads to the formation of a buffer gas flow, which can drag the molecules out of the cell \cite{Patterson2007}. As this drag hinders the free diffusive random walk of the molecules through the cell, and adds a directed motion towards the output nozzle, this effect can increase the output probability of the molecules. As this effect is omitted in the simulation, it predicts a continuously decreasing molecule flux for increasing buffer gas density, as shown in the figure. At this point, the simulation cannot be used as good approximation to the experiment any more. Therefore, these points are excluded when the theory is fitted to the experimental data to extract an effective collision cross section.

The fit parameter obtained from fitting the simulation to the experiment can be interpreted as an effective collision cross section. The resulting values are summarized in Table \ref{tab:CrossSections}. When comparing the curves and the corresponding fits for different temperatures in figure \ref{fig:FluxComparison}, one notices that for the measurement at $7\,$K, it takes a higher buffer gas density to reduce the count rate compared to the higher temperature measurement. However, the collision cross sections from the fits are almost identical. For the $30\,$K measurement, we obtain a value of $2.9\times10^{-15}\,$cm$^{2}$, while the $7\,$K measurement yields $3.0\times10^{-15}\,$cm$^{2}$. The statistical uncertainty is about $4\%$ in both cases. The shifted curve is a consequence of the changed collision dynamics. It is important to stress that these cross sections are based on a homogeneous scattering angle distribution. Hence, they cannot be directly compared to different calculations that are based on more realistic models.

Apart from the systematic error from the simulation, the experimental data contains systematic errors mainly from imprecise buffer gas density calibration inside the cell and from changes in experimental conditions between individual measurements. All in all, this adds up to an estimated common mode error of about $50\%$ and a variation between measurements of as much as $10\%$. Furthermore, as the collision energy changes during the cooling, for the interpretation of $\sigma$ it should be taken into account that the fit can only obtain an energy averaged collision cross section.

\renewcommand{\arraystretch}{1.2}
\begin{table}
	\begin{tabular}{| l | c |}
    \hline
    \multicolumn{1}{ |c| }{Settings} & Effective Collision Cross Section $\sigma$\\ \hline
    CH$_3$F + He, $7 $K & $3.0\cdot 10^{-15}\,$cm$^2$\\ \hline
		CH$_3$F + He, $30 $K & $2.9\cdot 10^{-15}\,$cm$^2$\\ \hline
		ND$_3$ + He, $8 $K & $2.2\cdot 10^{-15}\,$cm$^2$\\ \hline
		ND$_3$ + He, $47 $K & $8.9\cdot 10^{-16}\,$cm$^2$\\ \hline
		CH$_3$F + Ne, $30 $K & $2.4\cdot 10^{-15}\,$cm$^2$\\
    \hline
  \end{tabular}
	\caption{Energy averaged collision cross sections $\sigma$ for different molecules, buffer gases and temperatures}
  \label{tab:CrossSections}
\end{table}

\subsection{Benchmark exit velocity of molecules}
Another important benchmark for the simulations are the velocity distributions that can be derived from it. To check this, we use a bent quadrupole guide with a radius of $20\,$cm installed behind the cell nozzle to separate the molecule beam from the helium, and measure the velocity distribution by time-of-flight (TOF). The velocity distribution, $D^\prime (v)$, measured in this way corresponds to the one at the end of the bent quadrupole guide, for those molecules that survive the guiding process. The velocity distribution of interest for this work is, however, the one at the entrance of the guide, $D(v)$. $D(v)$ includes the collision effect at the vicinity of the cell output (i.e. the boosting process), but is independent from the quadrupole guiding process. To deduce $D(v)$, we use Monte Carlo trajectory simulations to produce an accurate $v$-dependent guiding efficiency of the bent quadrupole guide, $P(v)$, which leads to $D(v)=D^\prime(v)/P(v)$. A detailed description of these guiding efficiency simulations and their validation is given in \cite{Wu2016,WuPhD}.

\begin{figure}
\centering
  \includegraphics{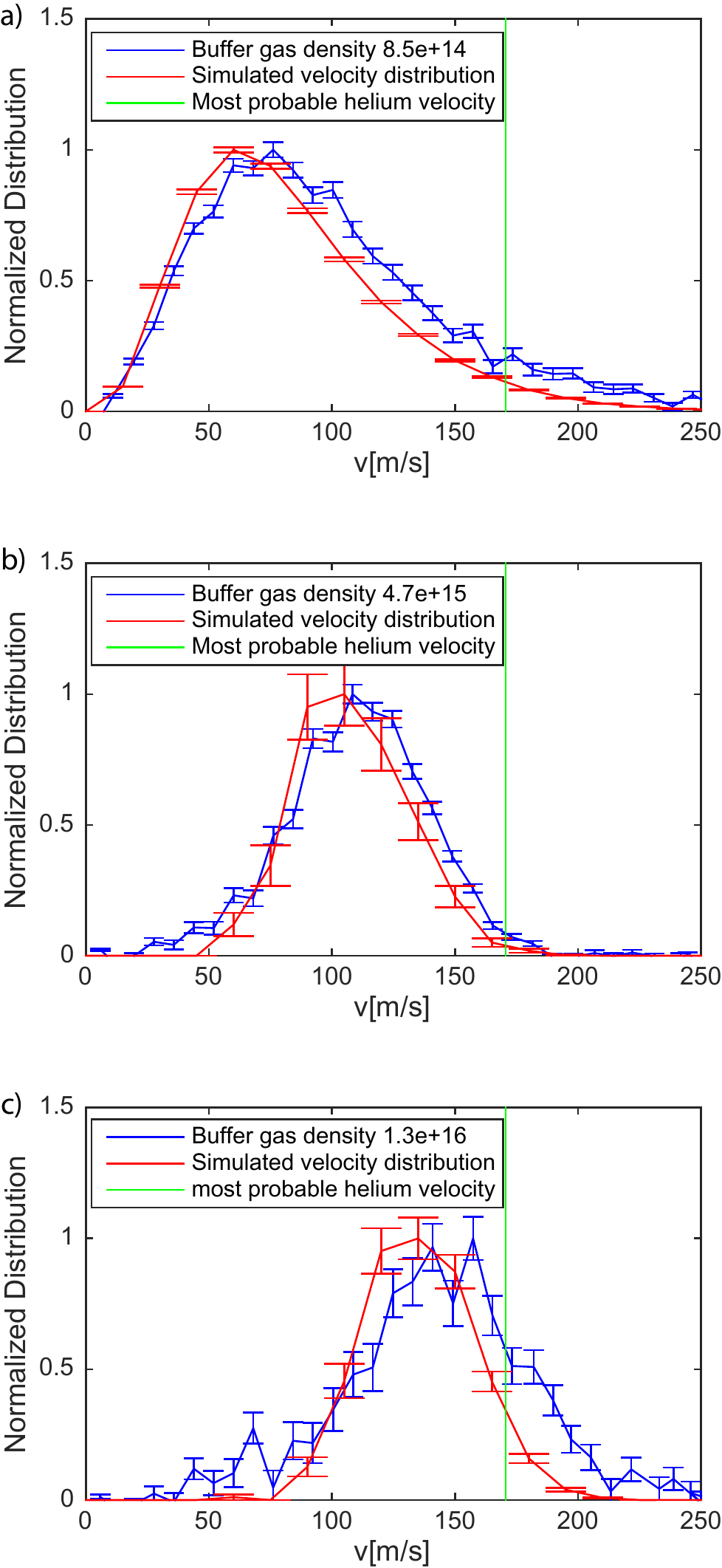}
  \caption{Comparison of velocity distributions obtained by simulations and experiment, in red and blue, respectively. a)-c) shows the results for an increasing buffer gas density. The green line shows the most probable velocity of buffer gas atoms in a Maxwell-Boltzmann distribution at cell temperature. The simulation fits the shape of the velocity distribution over a range of more than one order of magnitude in buffer gas density. }
  \label{fig:VelocityDistributions}
\end{figure}

Figure \ref{fig:VelocityDistributions} shows the comparison between the experimental data in blue and the simulated velocity distributions of the molecules in red. Additionally, the most probable velocity of  buffer gas atoms at cell temperature is given in green. The buffer gas density in the cell increases from a) to c), and is $8.5\times 10^{14}\,$cm$^{-3}$, $4.7\times 10^{15}\,$cm$^{-3}$, and $1.3\times 10^{16}\,$cm$^{-3}$, respectively. For the lowest buffer gas density, the velocity distribution does not show signs of boosting, and slow molecules even below $20\,$m/s are still present. For the higher densities, the slowest molecules are gone, and faster molecules up to $200\,$m/s appear. As soon as they approach the most likely helium atom velocity, the molecules cannot be boosted any further in the simulation. In the experiment, at these buffer gas densities, hydrodynamic effects would lead to a further acceleration of the molecules. At this point, the validity of the simulation breaks down.

To achieve the fit as shown in figure \ref{fig:VelocityDistributions}, the mean free path in the simulation has to be adjusted compared to the ones in the cell throughput studies. This fit results in a collision cross section of about $1.4\times 10^{-14}\,$cm$^2$, different compared to the $3.0\times 10^{-15}\,$cm$^2$ as given in table \ref{tab:CrossSections}. The reason for this disagreement is not fully understood. We suggest that it might result from the fact that the simulation uses a simple, hard sphere collision model, and a homogeneous scattering angle. The mismatch between the model and reality is captured by the fitted effective total collision cross section, and possibly the simple model cannot capture both effects. However, a single free fit parameter is enough to achieve a fit for the velocity distributions obtained at the different buffer gas densities. A more detailed investigation of the collisions, taking into account scattering angle distributions and inelastic collisions, might be able to remove this discrepancy, but is beyond the scope of this paper.

\section{Application to optimizing cell geometry}
We have shown that the simulation is capable of reproducing the number of molecules coming out of the cell, as well as their velocity distribution. Based on this, we can try to not only mirror the empirical data, but also to predict a result. Thereby, we propose a way to optimize the buffer gas cell that can save time compared to a purely empirical trial-and-error approach.  The implementation of this method we describe here is tailored for our experiment; however, it can be easily adapted to other situations. The figure of merit for us is the number of molecules leaving the cell at a velocity smaller than a threshold velocity $v_t$, so they can be guided by an electrostatic quadrupole guide. Furthermore, the molecules need to undergo enough collisions in the cell to efficiently thermalize with the buffer gas. For this example, the parameter we want to optimize is the length of the buffer gas cell. The exit nozzle diameter is fixed, as it has to match the size of the electric quadrupole guide we use in our experiments. The cell diameter will also be fixed to $3.5\,$cm, as before.

To find the optimum, the cell length $L$ is varied in the simulation, both smaller and larger than the original cell. The values used are $0.5\,$cm, $1\,$cm, $1.5\,$cm, $3\,$cm, and $5\,$cm. For these five simulation runs, we can compare the total flux of molecules and average number of collisions between molecules and buffer gas atoms, as they depend on buffer gas density and cell length. It is not necessary to obtain as many collisions as possible, but only a threshold number has to be reached for thermalization. Hence, the simulation should reveal which cell length/buffer gas density combination results in the highest possible molecule flux for a certain number of collisions, while avoiding boosting.

\begin{figure}
\centering
  \includegraphics[width=1\textwidth]{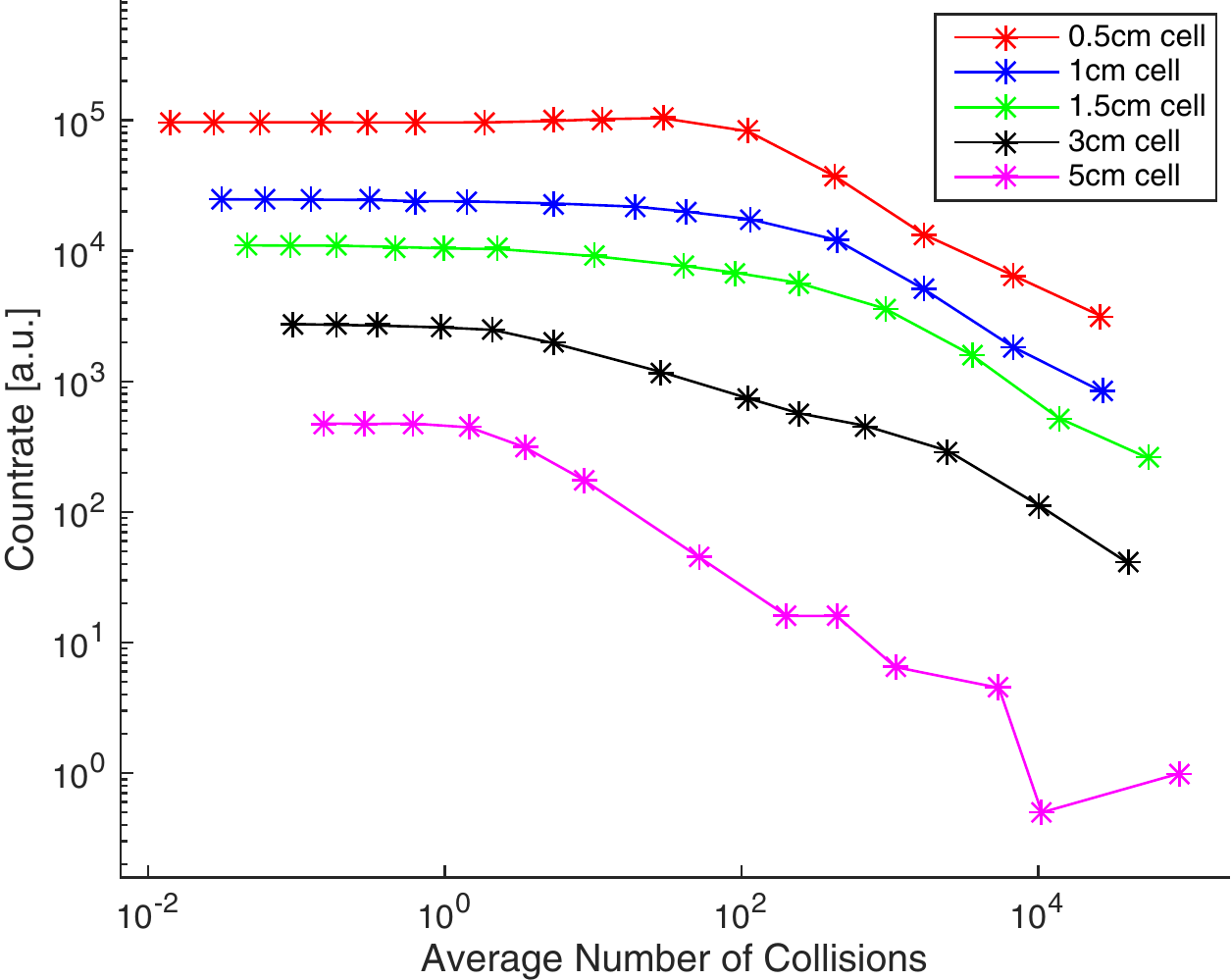}
  \caption{Number of molecules emerging from the buffer gas cell versus the average number of collisions per molecule in the cell, for different cell lengths.}
  \label{fig:CellLengthOptimization1}
\end{figure}

\begin{figure}
  \includegraphics[width=\columnwidth]{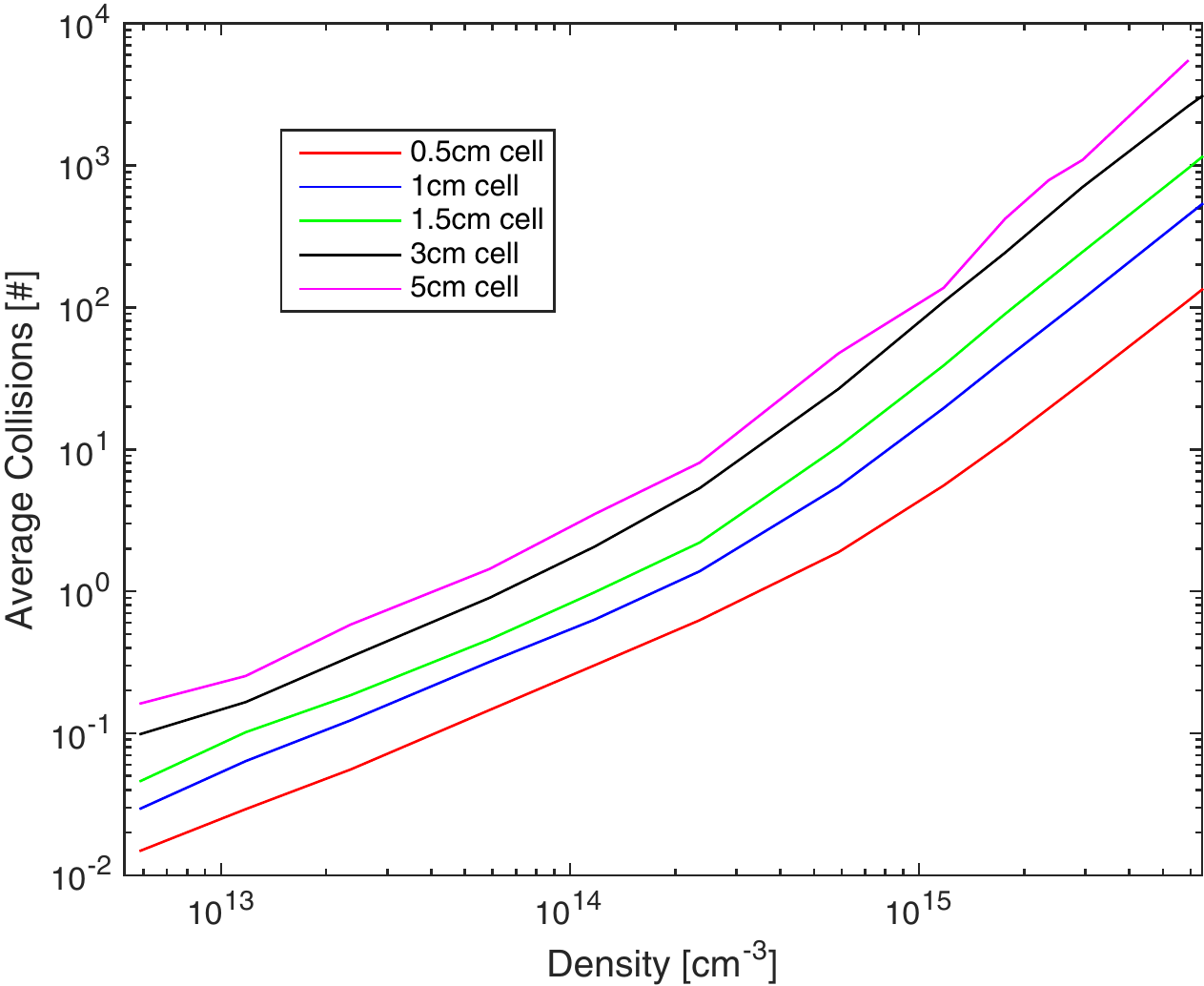}
  \caption{Average number of collisions per molecule versus the buffer gas density in the cell for various buffer gas cell lengths. }
  \label{fig:CellLengthOptimization2}
\end{figure}

In figures \ref{fig:CellLengthOptimization1} and \ref{fig:CellLengthOptimization2}, the output of molecules is plotted against the average number of collisions per molecule (\ref{fig:CellLengthOptimization1}), and the average collision number against the buffer gas density (\ref{fig:CellLengthOptimization2}), both for the various cell lengths. For all the plots, the test molecule is $CH_3F$, the buffer gas is helium, and the cell temperature is fixed to $7\,$K. 

It is apparent that a smaller cell delivers more molecules, as naively expected. For a fixed low buffer gas density, the molecule flux is inversely proportional to $L^2$. In the limit of no buffer gas and therefore no collisions, the cell acts as a simple aperture, and the acceptance angle scales inversely with $L^2$. In the intermediate and high density regime, the full Monte-Carlo simulation is necessary to achieve realistic results.

The simulation unsurprisingly shows (figure \ref{fig:CellLengthOptimization2}) that a higher buffer gas density must be applied for shorter cells to maintain the same number of collisions. Therefore, a shorter cell can always be compensated by a higher buffer gas density.  However, there is an upper limit for the buffer gas density, and therefore a lower bound for the cell length. It stems from the boosting effect of the buffer gas on the molecules for high buffer gas densities that accelerates molecules to a velocity bigger than $v_t$. This is problematic for the purpose of guiding molecules. From previous experiments \cite{Wu2016} we know, that the buffer gas density needs to remain below about $2 \cdot 10^{15}\,$cm$^{-3}$ for our system. The two requirements to reach a minimum number of about 20 collisions per molecule which is necessary for thermalization and to keep the buffer gas density below a threshold, are enough to predict an optimal cell length for our system between $0.5\,$cm and $1\,$cm, shorter than what was used for previous experiments. With a similar procedure, different parameters, like nozzle diameter, input or output nozzle positions, cell shapes, different molecule or buffer gas species, can be optimized. Moreover, the cell could be optimized for different figures of merit, like higher number of collisions for vibrational thermalization, molecule flux in a small solid angle, or maximum flux of very slow molecules to possibly circumvent the necessity of deceleration methods.

The simulations could not only reproduce the escape probability of the molecules and their velocity distributions. In the optimization process, they also revealed interesting, simple scaling laws, which help to qualitatively understand the buffer gas cell. For example, in the low buffer gas density regime, the number of collisions approximately depends on $\propto L\cdot n$. For higher densities, however, it is closer to $\propto L^2\cdot n^2$. The transition roughly happens when the process goes from thermalization of the molecules to a purely random walk like problem, after more than about 10 collisions appeared. The effect of the cell diameter was also briefly examined. Here, it was important to make the transverse direction of the cell large enough so that thermalization can occur, and a free diffusion process is possible. Therefore, it is important to make the cell radius larger than the cell length. With this condition fulfilled, the transverse size of the cell has a minor influence on the resulting molecule beam. With the qualitative understanding gained for these scalings, an improvement of the cell becomes easier.

\section{Summary and outlook}
In conclusion, the molecule propagation through a buffer gas cell was simulated, and it was shown that these simulations can accurately reproduce the buffer gas density dependent flux of molecules through the cell. Comparison to experimental data shows good agreement with a single fit parameter. Moreover, the empirically obtained velocity distributions of molecules in a quadrupole guide could be reproduced. The calculations do not depend on specific molecule properties other than mass, and can therefore easily be implemented for different molecules, buffer gases, temperatures or cell geometries. As all parameters of the simulated cell can easily be changed and computing times are relatively short even when performed on a standard office PC, different parameters can be easily tested before performing an experiment. 

With these results, we dramatically enhanced our qualitative understanding of the cryogenic buffer gas cell in the effusive regime. The analytical picture based on the Gambler's Ruin problem provides an intuition for the molecule dynamics inside the buffer gas environment. The detailed trajectory simulations give a more realistic picture of the movement of the molecules inside the cell, and moreover about the boosting process inside and outside the cell nozzle. Understanding the influence of the different parameters in the cell design can lead to more efficient cryogenic cells, which can be tailored to their specific tasks. 

In the future, the presented simulations could not only be used for the setup and optimization of buffer gas cells, but open a pathway to collision cross section measurements, for example on molecules interesting in astrophysics. At the moment, previously measured or calculated cross sections are inconclusive, as these numbers vary by more than an order of magnitude between $1\times10^{-15}\,$cm$^{2}$ and $2\times10^{-14}\,$cm$^{2}$ \cite{Broquier1988, Beaky1995, Loreau2015}. Our values fall within the same range, but previously mentioned systematic effects hinder a detailed examination of energy dependent molecule-atom collision cross sections. Nevertheless, a dedicated setup would be able to overcome these technical drawbacks. Buffer gas density calibration issues could be eliminated by implementing additional temperature and pressure gauges. The collision energy change can be minimized by using a double cell structure \cite{Patterson2007} to precool molecules, decreasing the energy loss of molecules inside the cell. Moreover, a more realistic collision model, for example based on the Eikonal approximation \cite{Sakurai2014}, could be realized.\\

\bibliographystyle{unsrt}
\bibliography{jobname}

\end{document}